\documentclass[12pt,preprint]{aastex}

\begin{document}

\title{Diffusion of Cosmic Rays in Expanding Universe. (I)}
\author{V. Berezinsky}
\affil{INFN - Laboratori Nazionali del Gran Sasso, I--67010
Assergi (AQ), Italy; \\
Institute for Nuclear Research of RAS, 60th October Revolution
prospect 7a,\\ 117312 Moscow, Russia}
\author{A. Z. Gazizov}
\affil{B.I. Stepanov Institute of Physics, Independence Avenue
68,\\ BY-220072 Minsk, Republic of Belarus}

\begin{abstract}
We present an analytic solution to diffusion equation for  high
energy cosmic rays in the expanding universe. The particles are
assumed to be ultra-relativistic and they can have energy losses
arbitrarily dependent on energy and time. The obtained solution
generalizes the Syrovatsky solution, valid for the case
when energy losses and diffusion coefficient are time-independent.
\end{abstract}

\keywords{extragalactic cosmic rays, diffusive propagation of cosmic rays.}

\section{Introduction}
\label{introduction} The cosmic rays is important component of
extragalactic medium, which is responsible for heating of
extragalactic gas and for production of various types
of radiation in the space. The charged cosmic ray particles
(electrons, protons and nuclei) propagate diffusively in
extragalactic space due to scattering in the magnetic fields. The
regime of diffusive propagation is reached if scattering
length, $\ell_{\rm sc}$, is much
less than the size of the considered region.

Typically, diffusion takes place in the magnetized
plasma, where the particle deflections occur due to scattering off
the turbulent pulsation and hydromagnetic waves. This process proceeds
in the resonant regime (see \citep{Lifshi}, Sections 55 and 61),
when a particle giro-radius is equal to a wave length. The
calculation of the diffusion coefficient for the resonant
scattering of particles off the hydromagnetic waves is presented
in the book by \cite{book}.

The value of diffusion coefficient is quite different for three
distinct regions in extragalactic space: voids, filaments and
clusters of galaxies. The diffusion in these regions
occurs only at energies when the diffusion
length, $\ell_{\rm diff}$,  is much smaller than
the size of the region $L$. In the opposite extreme case
particles propagate (quasi)rectilinearly.

For propagation of high energy particles from a single source at
point $\vec{r}_g$
the diffusion equation reads
\begin{equation}
\label{diff-eq-Sy} \frac{\partial}{\partial t}n_p(E,\vec{r},t) -
{\rm div} \left [ D(E,\vec{r},t)\nabla n_p\right]
-\frac{\partial}{\partial E}\left [ b(E,\vec{r},t)n_p\right ] =
Q(E,\vec{r},t)\delta^3(\vec{r}-\vec{r}_g) ,
\end{equation}
where $n_p(E,\vec{r},t)$ is the space density of particles $p$
with energy $E$ at time $t$ and at the point $\vec{r}$,
$D(E,\vec{r},t)$ is the diffusion coefficient,
$b(E,\vec{r},t)=-dE/dt$ describes the continuous energy
losses, and $Q(E,\vec{r},t)$ is the source generation function.

In the case when $D$, $b$ and $Q$ depend only on energy, the
method of exact analytic solution to the diffusion equation has
been suggested by Syrovatsky \citep{Syrov}. For a
single-source diffusion equation (\ref{diff-eq-Sy}) the solution
for spherically-symmetric case can be presented \citep{Bere90b} as
\begin{equation}
\label{syr-sol}
n_p(E,r)= \frac{1}{b(E)} \int_{E}^{\infty} dE_g Q(E_g) \frac{exp\left
[-\frac{r^2}{4\lambda(E,E_g)} \right ]} {\left [
4\pi\lambda(E,E_g)\right ]^{3/2}},
\end{equation}
where
\begin{equation}
\label{lambda-syr} \lambda(E,E_g) = \int_{E}^{E_g} d\varepsilon
\frac{D(\varepsilon)}{b(\varepsilon)}
\end{equation}
is the Syrovatsky variable which has the meaning of the
squared distance traversed by a particle in the observer
direction, while its energy diminishes from $E_g$ to $E$.

The other Syrovatsky variable,
\begin{equation}
\tau(E,E_g) = \int_{E}^{E_g} \frac{d\varepsilon}{b(\varepsilon)}
\label{tau-syr}
\end{equation}
has a meaning of time, during which the particle energy diminishes
from $E_g$ to $E$.

In this paper we shall present an analytic solution to
the diffusion equation in the expanding universe for the
case when $D$, $b$ and $Q$ are arbitrary functions of energy and
time. We use the computation method which differs
from that of \citep{Syrov}. While Syrovatsky used method
of the Green functions, we solve the equation for a single source.
Its position is described by the delta function
$\delta^3(\vec{r}-\vec{r}_g)$, and thus the equation for the
Fourier component does not contain $\delta$-functions.
This is the first order linear equation with partial
derivatives, which can be solved by a standard method,
introducing auxiliary characteristic equation.

As far as physical applications are concerned, the solution for a
single source is most general, because all other cases, e.g. with
homogeneously distributed sources or with single non-stationary source
(see \citep{Bere90b}) can be straightforwardly obtained from it.
\section{Diffusion equation in expanding universe}
\label{diff-eq} We shall use the
Friedmann-Robertson-Walker metric for the flat space and
radial direction, following \cite{Weinberg}
\begin{equation}
\label{metric}
ds^2=c^2 dt^2 - a^2(t) \vec{dx}^2= -
g_{\mu\nu}dx^{\mu\nu},
\end{equation}
where $diag\; g_{\mu\nu}=(-1,a^2,a^2,a^2)$ and $diag\; g^{\mu\nu}=
(-1,1/a^2,1/a^2,1/a^2)$, $\vec{x}$ is the spatial coordinate,
corresponding to comoving distance, and $a(t)$ is the scaling
factor of expanding universe, normalized as $a(t_0)=1$ at present
age of the universe $t_0$. The redshift $z$ is given by
$1+z=1/a(t)$ and $dt/dz$ by
\begin{equation}
-\frac{dt}{dz}=\frac{1}{H_0 (1+z) \sqrt{\Omega_m(1+z)^3+\Lambda}},
\label{dt/dz}
\end{equation}
where $H_0$ is the Hubble parameter at $z=0$ and $\Omega_m$ and
$\Lambda$ are cosmological mass density and vacuum energy in units
of the critical density.

The physical and proper distances are locally determined as 
$\vec{dr}=a(t)\vec{dx}$. For the {\em proper} distance it is assumed that 
$a(t)$ is not changed in the process of distance measurement between  
$\vec{x}=0$ and $\vec{x}$, and thus the proper distance between these two 
coordinates is $\vec{r}_{\rm prop}=a(t)\vec{x}$ , and velocity of the
universe expansion is $\vec{u}=\dot{a}\vec{x}=H(t)\vec{r}_{\rm prop}$.

For the {\em physical} distance it is assumed that measurement is
performed with help of the light signal ($ds^2=c^2 dt^2-a^2(t)dx^2=0$)
and the distance between the object with redshift $z$ and observer
with $z=0$ is 
\begin{equation}
r_{\rm ph}=\int_0^x a(t)dx=c\int_0^z dz \left |\frac{dt}{dz} \right |=
c\int_0^z \frac{dz}{H_0(1+z)\sqrt{\Omega_m (1+z)^3 +\Lambda}} .
\label{r_ph}
\end{equation}

Following \cite{Peebles} we shall use in this
paper the proper distance, denoting it by $\vec{r}$ and $r$ without
subscript.  Being the formal quantity, the proper distance coincides 
locally with the physical distance, and in most applications we
shall in fact use only local properties of the proper distance.
The positions of the
particles will be described by both proper distances $\vec{r}$
and by comoving distances $\vec{x}$, the latter can be considered as
the set of fixed coordinates in the expanding universe.

Expansion of the universe affects diffusion, and we shall take it
into account using the proper formalism.

The local observer sees the particle flux density produced by diffusion
\begin{equation}
j_k= - D \frac{\partial}{\partial x^k} n(\vec{x},t) .
\label{j_k}
\end{equation}
In case of isotropic diffusion (the diffusion coefficient is rotation
invariant),
$j_k$ is the covariant space vector, which together with proper space density
of particles $n$ forms the covariant 4-vector $j_{\mu}(\vec{x},t)=(n,j_k)$
(the Latin indices run through 1 - 3, and the Greek ones through 0 - 3).

The conservation of current $j^{\mu}$ can be written as \citep{Weinberg}
\begin{equation}
\frac{\partial}{\partial x^{\mu}} \left ( \sqrt{g} j^{\mu} \right )= 0,
\label{curr-conserv}
\end{equation}
where $g= |det~ g_{\mu\nu}|$ and $\sqrt{g}=a^3(t)$.
Differentiating Eq.~(\ref{curr-conserv}) with using
definition of the Hubble parameter $H(t)=\dot{a}(t)/a(t)$,
then transforming the contravariant component $j^k$ into covariant
one as $j^k=g^{km}j_m$, with $diag\; g^{km}= 1/a^2(t)$,
and finally using Eq.~(\ref{j_k}) for
\begin{equation}
\frac{\partial}{\partial x^k}j^k=-Dg^{ik}\frac{\partial}{\partial x^i}
\frac{\partial}{\partial x^k} n(\vec{x},t)= 
-\frac{D}{a^2}\nabla_x^2 n(\vec{x},t),
\label{curr-conserv1}
\end{equation}
one obtains the diffusion equation
\begin{equation}
\frac{\partial}{\partial t} n(\vec{x},t) +3 H(t) n(\vec{x},t) 
-\frac{D}{a^2} \nabla_x^2 n(\vec{x},t) =0,
\label{diff-eq-short}
\end{equation}
where $3H(t)n$ term describes expansion of the universe.

The derived equation can be obtained from conservation equation (9.15)
of \cite{Peebles} excluding there the advection term and adding the 
diffusion term.  

The derivation of the diffusion equation above is rather formal, and
now we shall present it in the transparent and physically clear way. 

We shall operate here and everywhere below with particle density $n$
in the expanding
volume using two sets of the coordinates ($\vec{r},t$) and
($\vec{x},t$). This density is given by 
\begin{equation}
n(\vec{r},t)=n(\vec{x},t), 
\label{n}
\end{equation}
with $\vec{r}(t)=a(t)\vec{x}$. Differentiating $n(\vec{r},t)$, 
\begin{equation}
\frac{d}{dt} n(\vec{r},t)=\left (\frac{\partial n}{\partial t} \right )_r
+ \frac {\partial n}{\partial \vec{r}}\frac{d\vec{r}}{dt}=
\left (\frac{\partial n}{\partial t} \right )_r + 
H(t) \vec{r}\; \frac{\partial n}{\partial \vec{r}} ,
\label{dn/dt}
\end{equation}
and using $d n(\vec{r},t)/dt=\partial n(\vec{x},t)/\partial t$ from (\ref{n})
one obtains the Peebles relation given by Eq.~(9.13) from \citep{Peebles}:
\begin{equation}
\left (\frac{\partial n}{\partial t} \right )_x=
\left (\frac{\partial n}{\partial t} \right )_r + 
H(t) \vec{r}\; \vec{\nabla}_r n ,
\label{peebl}
\end{equation}
where subscripts $x$ and $r$ indicate basis $(\vec{x},t)$ and 
$(\vec{r},t)$, respectively. 

We shall obtain the diffusion equation from conservation of number of
particles. Consider a sphere of radius $x$, which expands in the basis 
$(\vec{r},t)$ as $r(t)= a(t)x$. The number of particles inside this
sphere is changing only due to diffusive flux, which is 
defined as $\vec{j}=-D(\vec{r},t)\vec{\nabla}_r n(\vec{r},t)$. The
corresponding equation reads:
\begin{equation}
\frac{d}{dt}\int_{V(t)} dV n(\vec{r},t)= -\int_{S(t)} \vec{j}d\vec{s}= 
\int_{V(t)} dV div [D(\vec{r},t)\vec{\nabla}_r n(\vec{r},t)],
\label{gauss}
\end{equation}
where $S(t)$ is the expanding sphere and $V(t)$ is the volume inside
it. In Eq.~(\ref{gauss}) the Gauss theorem was used. 

Performing differentiation in
the lhs of Eq.~(\ref{gauss}) with respect to time, and
taking into account expanding of the elemental volume $dV$ with time,   
$$
\frac{d}{dt}\delta V=\frac{d}{dt}\left [ a^3(t)(\delta V)_{\rm comov}\right ]
=3H(t) \delta V,
$$
and using for $dn/dt$ Eq.~(\ref{dn/dt}), one obtains
\begin{equation}
\left (\frac{\partial n}{\partial t}\right )_r  + 
H(t) \vec{r}\;\vec{\nabla}_r n + 3 H(t) n - 
div \left [ D(r,t) \vec{\nabla}_r n(\vec{r},t) \right ] =0.
\label{diff-eq-r}
\end{equation}
This is the diffusion equation in the physical basis $(\vec{r},t)$. 
One may add to the rhs the source term $Q_0\delta^3(\vec{r}-\vec{r}_g)$. 

The sum of the second and third terms in 
Eq.~(\ref{diff-eq-r}) merges into $\vec{\nabla}_r(n\vec{u})$,
where in our case $\vec{u}=H(t)\vec{r}$~ is expansion velocity 
of the universe, and in
absence of the diffusion term we obtain Eq.~(9.11) from \citep{Peebles}. 

In the basis $(\vec{x},t)$ the first two terms in 
Eq.~({\ref{diff-eq-r}) give $\partial n(\vec{x},t)/\partial t$ 
(see Eq.~\ref{peebl}), and we arrive at the diffusion equation 
in the form 
\begin{equation}
\left (\frac{\partial n}{\partial t} \right )_x + 3 H(t) n(\vec{x},t) 
-\frac{D(t)}{a^2(t)} \nabla_x^2 n(\vec{x},t) = 
\frac{Q_0}{a^3(t)}\delta^3(\vec{x}-\vec{x}_g), 
\label{diff-eq-x}
\end{equation}
in accordance with Eq.~(\ref{diff-eq-short}). Deriving Eq.~(\ref{diff-eq-x})
we assumed that $D$ is a function of time only. 

Eqs.~(\ref{diff-eq-short}) and (\ref{diff-eq-x}) are the
simplified diffusion equations  without energy loss and full  source
terms, which should be included additionally. Introducing these
terms as $-\frac{\partial}{\partial E}(nb)$ and
$Q(E,t)\delta^3(\vec{r}-\vec{r_g})$ we arrive 
at the diffusion
equation for expanding universe in the  basis $(\vec{x},t)$ convenient 
for its solution:
\begin{equation}
\frac{\partial n}{\partial t} - b(E,t)\frac{\partial n}
{\partial E}+3H(t)n - n\frac{\partial b(E,t)}{\partial E}
-\frac{D(E,t)}{a^2(t)}\nabla_x^2 n = \frac{Q(E,t)}{a^3(t)}
\delta^3(\vec{x}-\vec{x}_g),
\label{diff-basic-x}
\end{equation}
where the total  energy losses $dE/dt= -b(E,t)$ may be presented as 
a sum of collisional energy losses $b_{int}(E,t)$ and adiabatic
energy losses $H(t)E$, and $Q(E,t)$ is the number of particles with 
energy $E$ produced at time $t$ per unit time. 
In Eq.~(\ref{diff-basic-x}) $b(E,t)$ and   
$D(E,t)$ are the arbitrary functions of time and energy, and 
$n=n(t,\vec{x},p)$ must depend in fact on $\vec{x}-\vec{x}_g$.

Note, that $n(t,\vec{x},p)$ in Eq.~(\ref{diff-basic-x}) is not the
distribution function $f(t,\vec{x},p)$, which is defined in
statistical physics (see e.g. \citep{Lifshi}) as the
density in phase space. The relation between them is given by
$n(t,\vec{x},p)=4\pi p^2 f(t,\vec{x},p)$. We remind that here is
considered the ultra-relativistic case $p \approx E/c$.

\section{Analytic solution to the diffusion equation}
\label{anal-solution} We shall introduce the Fourier
transformation
\begin{equation}
\label{F-trans}
n(t,\vec{x},E)=\frac{1}{(2\pi)^3}\int
d\vec{\omega} f_{\omega}(E,t)
e^{i\vec{\omega}(\vec{x}-\vec{x}_g)} .
\end{equation}
Using the Fourier expansion of the $\delta$ function
\begin{equation}
\delta^3(\vec{x}-\vec{x}_g)=\frac{1}{(2\pi)^3}\int d\vec{\omega}
e^{i\vec{\omega}(\vec{x}-\vec{x}_g)},
\label{F-delta}
\end{equation}
we obtain from Eq.~(\ref{diff-basic-x}) equation for the Fourier
components $f_{\omega}(E,t)$:
\begin{equation}
\label{F-eq}
\frac{\partial}{\partial t}f_{\omega}(E,t)
-b(E,t)\frac{\partial}{\partial E}f_{\omega}(E,t) +\left [
3H(t)-\frac{\partial b(E,t)}{\partial E}+\vec{\omega}^2
\frac{D(E,t)}{a^2(t)}\right ]
f_{\omega}(E,t)=\frac{Q(E,t)}{a^3(t)}.
\end{equation}
The characteristic equation for Eq.~(\ref{F-eq}) is
\begin{equation}
\label{char}
\frac{dE}{dt}= -b(E,t),
\end{equation}
with the solution $\mathcal{E}'=E'(E,t,t')$ being identical with
the generation-energy trajectory $E_g(E,t,t')$ used in \citep{Bere88} 
and \citep{Bere02}, where $E_g$ is the
energy with which a particle must be generated at time $t'$ in
order to have at $t$ the observed energy $E$. Here and everywhere
below we shall use the energy $E$ at the time $t$ of observation to
mark a characteristic trajectory $\mathcal{E}'=E'(E,t,t')$. Sometimes for
brevity we shall omit $t$.

The solution to Eq.~(\ref{F-eq}) with energies taken on
characteristics is given by
\begin{equation}
\label{sol-F} f_{\omega}(E,t)=\int\limits_{t_g}^t
dt'\frac{Q(\mathcal{E}',t')}{a^3(t')} \exp \left \{-\int
\limits_{t'}^t dt''\left[ 3H(t'')-\frac{\partial
b(\mathcal{E}'',t'')} {\partial \mathcal{E}''} +\vec{\omega}^2
\frac{D(\mathcal{E}'',t'')}{a^2(t'')}\right ]\right\}
\end{equation}
where $t_g$ is the generation time.

We introduce now $\lambda(E,t,t')$, the analogue of the Syrovatsky
variable given by Eq.~(\ref{lambda-syr})
\begin{equation}
\label{lambda} \lambda (E,t')= \int_{t'}^t dt''
\frac{D(\mathcal{E}'',t'')}{a^2(t'')},
\end{equation}
where $\mathcal{E}''=E''(E,t,t'')$ is the characteristic
trajectory.

The exponent in Eq.~(\ref{sol-F}) can be simplified using the notation
\begin{eqnarray}
\alpha_{\omega}(E,t,t')&=& \int_{t'}^t dt''\left [ 3H(t'') -
\frac{\partial b(\mathcal{E}'',t'')}{\partial \mathcal{E}''}
\right ]+
\vec{\omega}^2\lambda (E,t') \nonumber\\
&=&\ln \left (\frac{1+z'}{1+z} \right )^3 + \vec{\omega}^2\lambda
(E,t') - \int_{t'}^t dt''\frac {\partial
b(\mathcal{E}'',t'')}{\partial \mathcal{E}''} \label{alpha}
\end{eqnarray}
Thus,
\begin{equation}
e^{-\alpha_{\omega}(E,t,t')}=\left (\frac{1+z}{1+z'} \right )^3
e^{-\vec{\omega}^2\lambda(E,t')}\exp\int_{t'}^t dt''
\frac{\partial b(\mathcal{E}'',t'')}{\partial \mathcal{E}''}.
\label{exp}
\end{equation}
Coming back from the Fourier component $f_{\omega}$ to density $n$, using
identity
$$
i\vec{\omega}(\vec{x}-\vec{x}_g)-\vec{\omega}^2\lambda=
-\lambda \left [\vec{\omega}-i\frac{\vec{x}-\vec{x}_g}{2\lambda}\right ]^2
-\frac{(\vec{x}-\vec{x}_g)^2}{4\lambda},
$$
and performing integration
\begin{equation}
\int d\vec{\omega}\exp \left [-\lambda\left ( \vec{\omega}
-i\frac{(\vec{x}-\vec{x}_g)^2}{2\lambda} \right )^2\right]=
(\pi/\lambda)^{3/2} ,
\end{equation}
we obtain the solution as
\begin{equation}
n(t,\vec{x},E)=\frac{\pi^{3/2}}{(2\pi)^3}\int_{t_g}^tdt'\frac{Q(\mathcal{E}',t')}{(1+z)^3}
\;\frac{\exp \left
[-\frac{(\vec{x}-\vec{x}_g)^2}{4\lambda(E,t')}\right ]}
{[\lambda(E,t')]^{3/2}}\;\exp\left[ \int_{t'}^t dt''
\frac{\partial b(\mathcal{E}'',t'')} {\partial
\mathcal{E}''}\right]
\end{equation}
Now we shall find the solution for $t=t_0$ ($z=0$) changing the variables
$t' \to z$, $t'' \to z'$ and presenting the energy loss as the sum of
adiabatic and collisional (interaction) energy losses:
\begin{equation}
\label{b(E,t)}
b(E,t)=H(t)E + b_{int}(E,t) .
\end{equation}
It gives
\begin{eqnarray}
n(t_0,\vec{x},E) &= &\int_0^{z_g} dz \left |\frac{dt}{dz} \right |
(1+z) Q(\mathcal{E}_g,z) \exp \left [\int_0^z dz'\left
|\frac{dt'}{dz'}\right | \frac{\partial
b_{int}(\mathcal{E}',z')}{\partial \mathcal{E}'}\right ]\times  \nonumber \\
& & \frac{\exp[-(\vec{x}-\vec{x}_g)^2/ 4\lambda(E,z)]}{[4\pi
\lambda(E,z)]^{3/2}}, \label{diff-sol}
\end{eqnarray}
where $\mathcal{E}_g=E_g(E,z)$ is the generation energy in the
source taken on the characteristic.

In this expression one can easily distinguish the term $dE_g/dE$,
given for the case of protons interacting with CMB in
\citep{Bere88} and \citep{Bere02}. In our case it equals to
\begin{equation}
\frac{dE_g}{dE}= (1+z)\exp\left ( \int_0^z dz'\left
|\frac{dt'}{dz'} \right | \frac{\partial
b_{int}(\mathcal{E}',z')}{\partial \mathcal{E}'} \right ) .
\label{dE_g/dE}
\end{equation}
Now the solution for the case of arbitrary energy losses
$b_{int}(E,z)$ can be written in the compact form
\begin{equation}
n(t_0,\vec{x},E)=\int_0^{z_g} dz \left |\frac{dt}{dz} \right |
Q[E_g(E,z),z]\;\frac{\exp[-(\vec{x}-\vec{x}_g)^2/4\lambda(E,z)]}
{[4\pi\lambda(E,z)]^{3/2}}\;\frac{dE_g}{dE}, 
\label{solut-gen}
\end{equation}
where $n(t_0,\vec{x},E)$ actually depends on $\vec{x}-\vec{x}_g$
as was foreseen above.

In the case of ultra-high energy protons interacting with CMB,
$\partial b_{int}/\partial E$ is given according to \cite{Bere88}
and \cite{Bere02} by
\begin{equation}
\frac{\partial b_{int}(E,t)}{\partial E}= (1+z)^3\left
[\frac{db_0(E')} {dE'}\right ]_{E'=(1+z)E_g(E,z)}
\end{equation}
and
\begin{equation}
\frac{dE_g}{dE}= (1+z)\exp\left [ \frac{1}{H_0} \int_0^z dz
\frac{(1+z)^2}{\sqrt{\Omega_m(1+z)^3+\Lambda}} \left (
\frac{db_0(E')}{dE'} \right )_{E'=(1+z)E_g(E,z)}\right ] ,
\label{dE_g/dE-cmb}
\end{equation}
where $b_0(E)$ is the interaction energy loss at $z=0$.

Eq.~(\ref{solut-gen}) presents our final result in the form of the
particle density $n(t_0,\vec{x},E)$ in terms of comoving distances
$\vec{x}$ for arbitrary energy loss $b_{int}(E,t)$ and diffusion
coefficient $D(E,t)$. The main difference with the
Syrovatsky solution is due to $\lambda (E,t)$ given by
Eq.~(\ref{lambda}) and $dE_g/dE$ given by Eq.~(\ref{dE_g/dE}).
\cite{Lemoine} has heuristically written a
solution to the diffusion equation in expanding universe,
including only adiabatic energy losses. He used the
Syrovatsky solution for the diffusion  equation in terms of conformal
time $\eta$. However, the Syrovatsky solution is not valid there,
because both diffusion coefficient and the Hubble parameter depend
on conformal time $\eta$. Besides, our equation
(\ref{diff-basic-x}) differs from the Lemoine's equation
by extra $a(t)$ in the diffusion term.

Eq.~(\ref{solut-gen}) becomes the Syrovatsky solution
(\ref{syr-sol}), when $D(E,t)$ and $b(E,t)$ do not depend on time
and $a(t)=1$. In this case $dE_g/dE=b(E_g)/b(E)$ \citep{Bere88},
and using $b(E_g)dt=dE_g$ one obtains Eq.~(\ref{syr-sol}) with
$$
\lambda= \int dt'D(E',t')=\int_E^{E_g}\frac{dE'}{b(E')}D(E') ,
$$
being the Syrovatsky variable (\ref{lambda-syr}). Proper
transition to the Syrovatsky solution is one of the tests of
our solution (\ref{solut-gen}).
\section{Tests of the method}
We want to test our method of equation solution for some cases,
when solutions are known. One of them is given by the case of
time-independent energy loss and diffusion coefficient, which must
result for $a(t)=1$ in the Syrovatsky solution. In 
Section \ref{anal-solution} it has been already demonstrated that
our solution passes this test.

The second test  consists in the convergence to the universal spectrum,
and the third, most important one, examines the cosmological part of
our solution: The solution of equation for the rectilinear propagation
is obtained by the same method as for diffusion and 
result coincides with well known formula.
Below we shall describe these tests.   
\subsection{Convergence to the universal spectrum}
According to the propagation theorem \citep{AloisioBere04}
in the case when distances between sources ($d$) become
smaller than propagation and  interaction lengths (e.g.\
when $d \to 0$), the diffuse spectrum has an universal
form, independent of the mode of propagation.

For the power-law generation spectrum with exponent $\gamma_g >2$ ,
\begin{equation}
Q(E,z)=(\gamma_g-2)L_0(1+z)^{\alpha} E^{-\gamma_g}, \label{Q}
\end{equation}
where $L_0$ is the particle luminosity of a source at
$z=0$ and $\alpha$ takes into account a hypothetical
cosmological evolution (in Eq.~(\ref{Q}) all energies are measured
in GeV, luminosity in GeV s$^{-1}$ and $E_{\rm min}=1$~GeV).
In this case the universal spectrum for the diffuse
flux $J(E)$ is given \citep{AloisioBere04} as
\begin{equation}
J(E)=\frac{c}{4\pi}{\cal L}_0(\gamma_g-2) \int_0^{z_{\rm max}} dz
\left |\frac{dt}{dz} \right |(1+z)^m  E_g^{-\gamma_g}(E,z)
\frac{dE_g}{dE} ,
\label{univ-sp}
\end{equation}
where ${\cal L}_0=L_0 n_s(0)$ is the emissivity at $z=0$,
$n_s(0)$ is density of the sources at $z=0$,
$n_s(z)=n_s(0)(1+z)^{\beta}$ describes the hypothetical source
evolution and $\alpha+\beta=m$.

Let us now calculate the diffuse flux from the density of the
particles $n(t_0,\vec{x},E)$ given by our general solution
(\ref{solut-gen}). According to the propagation theorem we must obtain
the universal spectrum (\ref{univ-sp}).

Assuming the homogeneous distribution of the sources $n_s(z)$ in the space
(which provides the propagation theorem) we can find the diffuse flux
multiplying $Q(E_g,z)$ in Eq.~(\ref{solut-gen}) by the density of the sources
$n_s(z)=n_s(1+z)^{\beta}$ and integrating over positions of the
sources in the coordinate space. Using (\ref{Q}) we have
\begin{equation}
J_p(E)=\frac{c}{4\pi}(\gamma_g-2){\cal L}_0\int 4\pi x_s^2
dx_s \int_0^{z_{\rm max}} dz \left |\frac{dt}{dz} \right |(1+z)^m
E_g^{-\gamma_g}(E,z)\frac{\exp [-x_s^2/4\lambda(E,z)]}
{[4\pi\lambda (E,z)]^{3/2}}\;
\frac{dE_g}{dE} ,
\label{J_p}
\end{equation}
where $\vec{x}_s = \vec{x}-\vec{x}_g$. Changing the order of
integration and using
\[
\int_0^{\infty}dx \frac{4\pi x^2}{(4\pi\lambda)^{3/2}}
\exp\left(-\frac{x^2}{4\lambda}\right)=1,
\]
we arrive indeed at the universal spectrum
(\ref{univ-sp}), as it must be. The assumption of the power-law
generation spectrum in Eq.~(\ref{Q}) does not reduce the
generality of the proof.
\subsection{Rectilinear propagation}
The third test which we shall study here is less trivial 
than other two and it
examines the cosmological part of our solution.  We consider
the rectilinear propagation of ultra-relativistic
particles. This case is well known, and is given by light
propagation. We shall find a solution for this case, solving the
propagation equation by the method employed in Section
\ref{anal-solution}.

The rectilinear propagation of particles with velocity $\vec{v}$ 
results in adding this velocity to the expansion velocity 
$H(t)\vec{r}$~ in the rhs of Eq.~(\ref{dn/dt}) and in excluding the diffusion 
term from the equation. Introducing the unit vector $\vec{e}$ in the 
direction of propagation, and using $\vec{dr}=a(t)\vec{dx}$ we obtain
from Eqs.~(\ref{dn/dt}) and  (\ref{diff-basic-x}):
\begin{equation}
\frac{\partial n}{\partial t} +
\frac{c\vec{e}}{a(t)}\frac{\partial n}{\partial\vec{x}}
- b(E,t)\frac{\partial n}{\partial E}+3H(t)n - n\frac{\partial b}{\partial E}
= \frac{Q(E,t)}{a^3(t)} \delta^3(\vec{x}-\vec{x}_g),
\label{rect-basic}
\end{equation}
where $n=n(t,\vec{x},E)$.

Performing the Fourier transformations (\ref{F-trans}) and (\ref{F-delta}),
and using
$$
\frac{c\vec{e}}{a(t)}\frac{\partial}{\partial \vec{x}}n(t,\vec{x},E)=
\frac{1}{(2\pi)^3}\int d\vec{\omega}\frac{i(\vec{e}\vec{\omega})c}{a(t)}
f_{\omega}(E,t)e^{i\vec{\omega} (\vec{x}-\vec{x}_g)},
$$
we obtain from (\ref{rect-basic}) the following equation for the
Fourier transform $f_{\omega}(E,t)$:
\begin{equation}
\frac{\partial}{\partial t}f_{\omega}(E,t) -
b(E,t)\frac{\partial}{\partial E}f_{\omega}(E,t)+ \left [
3H(t)-\frac{\partial b(E,t)}{\partial E}+
i\frac{(\vec{e}\vec{\omega})c}{a(t)}\right ] f_{\omega}(E,t)
=\frac{Q(E,t)}{a^3(t)} .
\label{F-eq-rect}
\end{equation}
With the help of the characteristic equation
(\ref{char}), which gives $\mathcal{E}'=E'(E,t,t')$, the solution of
Eq.~(\ref{F-eq-rect}) is found as
\begin{equation}
f_{\omega}(E,t)=\int_{t_g}^t dt'
\frac{Q[\mathcal{E}',t']}{a^3(t')} e^{-\alpha_{\omega}(E,t,t')},
\label{sol-F-rect}
\end{equation}
with
\begin{equation}
\alpha_{\omega}(E,t,t')=\int_{t'}^t dt'' \left [
3H(t'')-\frac{\partial b(\mathcal{E}'',t'')}{\partial
\mathcal{E}''}+ i\frac{(\vec{e}\vec{\omega})c}{a(t'')} \right ].
\label{alpha-rect}
\end{equation}
To calculate $\alpha_{\omega}(E,t,t')$ we
use
$$
\int_{t'}^t dt'' H(t'')=\ln \frac {1+z'}{1+z},
$$
$$
i(\vec{e}\vec{\omega})c\int_{t'}^t \frac{dt''}{a(t'')}=
i\vec{\omega}[\vec{x}(t)- \vec{x}(t')] ,
$$
and $b(E,t)=H(t) E+ b_{int}(E,t)$. Then it follows
\begin{equation}
e^{-\alpha_{\omega}(E,t,t')}=\left (\frac{1+z}{1+z'}\right )^2
e^{-i\vec{\omega}(\vec{x}-\vec{x}')}\exp \left [\int_z^{z'} dz''
\left |\frac{dt''}{dz''}\right | \frac{\partial
b_{int}(\mathcal{E}'',z'')}{\partial \mathcal{E}''}\right ].
\end{equation}
Coming back to $n(t,\vec{x},E)$ we have
\begin{eqnarray}
n(t,\vec{x},E) & = &\int \frac{d\vec{\omega}}{(2\pi)^3}
e^{i\vec{\omega}(\vec{x}-\vec{x}_g)} \int_{t_g}^t
dt'\frac{Q(\mathcal{E}',t')}{a^3(t')} \left
(\frac{1+z}{1+z'}\right
)^2 e^{i\vec{\omega}(\vec{x}'-\vec{x})} \times \nonumber \\
& & \exp \left [\int_z^{z'} dz''\left |\frac{dt''}{dz''}\right |
\frac{\partial b_{int}(\mathcal{E}'',z'')}{\partial \mathcal{E}''}
\right] . \label{sol-rect}
\end{eqnarray}
Using the Fourier expansion of $\delta$ function and its properties,
\[
\int \frac{d\vec{\omega}}{(2\pi)^3}e^{i\vec{\omega}(\vec{x}'-\vec{x}_g)}
=\delta^3 [\vec{x}(t')-\vec{x}_g]=
\frac{1}{4\pi x_g^2}\delta [x(t')-x_g] =
\frac{a(t_g)}{4\pi c x_g^2}\delta(t'-t_g) ,
\]
we get the solution at
$t=t_0$ (z=0) and $\vec{x}=0$
\begin{equation}
n(t_0,\vec{x}=0,E)= \int dt'\frac{Q(\mathcal{E}',t')}{a(t')}
\frac{a(t_g)}{4\pi c x_g^2}\delta (t'-t_g) \exp \left [
\int_0^{z'} dz''\left |\frac{dt''}{dz''}\right | \frac{\partial
b_{int}(\mathcal{E}'',z'')}{\partial \mathcal{E}''}\right ] ,
\end{equation}
where $a(t')(1+z')=1$ is used.

Performing integration over $t'$ and using Eq.~(\ref{dE_g/dE}) we
finally have
\begin{equation}
n(t_0,E)=\frac{Q(E_g,t_g)}{4\pi c x_g^2 (1+z_g)}\frac{dE_g}{dE} ,
\label{n(E)}
\end{equation}
where $x_g$ is the comoving distance to a source.

One can compare $n(t_0,E)$ from Eq.~(\ref{n(E)}) with energy flux
$F$ (in erg cm$^{-2}$s$^{-1}$) of photons emitted by a
source with luminosity $L$ at comoving distance $x_g$, as given by
Eq.~(2.42) in the book by \cite{KT}:
\begin{equation}
F=\frac{L}{4\pi x_g^2 (1+z_g)^2} ,
\label{KT}
\end{equation}
where one factor $(1+z_g)$ arises from the time dilation
and the other one from energy redshift.

In our case $L \to Q(E_g)dE_g$ and $F \to cn(E)dE$, with one
factor $(1+z_g)$ disappearing because we consider the number of
particles instead of the luminosity and energy flux.

Therefore, our method passes this test, too.

\section{Conclusions}
We shall conclude giving the formulae in the form convenient for
practical use.

Our basic diffusion equation (\ref{diff-basic-x}) for
ultra-relativistic particles propagating from a single source, reads
\begin{equation}
\frac{\partial n}{\partial t} - b(E,t)\frac{\partial n}
{\partial E}+3H(t)n - n\frac{\partial b(E,t)}{\partial E}
-\frac{D(E,t)}{a^2(t)}\nabla_x^2 n = \frac{Q(E,t)}{a^3(t)}
\delta^3(\vec{x}-\vec{x}_g),
\label{diff-basic}
\end{equation}
where $n(t,\vec{x},E)$ is the particle number density per
unit energy in expanding volume of the universe, $\vec{x}$ is
coordinate corresponding to the comoving distance, $dE/dt
=-b(E,t)$ describes the total energy losses, which include
adiabatic $H(t)E$ and interaction $b_{int}(E,t)$ energy losses,
and $Q(E,t)$ is the generation function, given by the
number of particles generated by a single source, located
at coordinate $\vec{x}_g$, per unit energy and unit time.

Solution of Eq.~(\ref{diff-basic}) can be presented in
the spherically-symmetric case as
\begin{equation}
n(x_g,E)=\int_0^{z_g} dz \left |\frac{dt}{dz} \right |
Q[E_g(E,z),z]\;\frac{\exp[-x_g^2/4\lambda(E,z)]}
{[4\pi\lambda(E,z)]^{3/2}}\;\frac{dE_g}{dE},
\label{solution}
\end{equation}
where
\begin{equation}
\lambda (E,z)= \int_0^z dz' \left |\frac{dt'}{dz'} \right |
\frac{D(\mathcal{E}',z')}{a^2(z')} , \label{lambda(E,z)}
\end{equation}
\begin{equation}
\frac{dE_g}{dE}= (1+z)\exp\left [ \int_0^z dz'\left
|\frac{dt'}{dz'} \right | \frac{\partial
b_{int}(\mathcal{E}',z')}{\partial \mathcal{E}'} \right ], 
\label{dEg/dE}
\end{equation}
where $\mathcal{E}'=E'(E,z')$ is a characteristic trajectory, which
gives energy $E'$ of a particle at epoch $z'$, if this
energy is $E$ at $z=0$; $E_g(E,z)$ has the same meaning. 
The upper limit $z_g$ in the integral of
Eq.~(\ref{solution}) is provided by maximum energy of acceleration
as $E_g(E,z_g)=E_{\rm max}$, or by $z_{\rm max}$, what
is smaller.

The solution (\ref{solution}) is intentionally presented in the
form similar to the Syrovatsky solution with $\lambda(E,z)$ being
an analogue of the Syrovatsky variable. However, this solution
cannot be obtained neither from the
Syrovatsky solution (\ref{syr-sol}) nor by the Syrovatsky method
of solution, which is based on time-independent quantities
$\lambda$ and $\tau$ as new variables.

Eq.~(\ref{solution}) is convenient for various applications.

In case the sources are distributed homogeneously with density
$n_s(x_g)$ the total density of particles $n_{\rm tot}(E)$ can be
found by integration of Eq.~(\ref{solution}) over $4\pi x_g^2
dx_g$, with the diffuse flux given by $J(E)=(c/4\pi)n_{\rm
tot}(E)$. In case of discrete distribution of the sources, the
density $n_{\rm tot}(E)$ must be found by summation over sources,
like it is done in \citep{AloisioBere04}. For non-stationary
source, being switched on at redshift $z_1$ and switched off at
redshift $z_2$, the limits of integration in Eq.~(\ref{solution})
must be taken as $z_1$ and $z_2$ with additional acceleration cut
on $z_{\rm max}$ given by equation $E_g(E,z_g) \leq E_{\rm max}$ 
(see also \cite{Bere90b}).

\section*{Acknowledgments}
We acknowledge participation of Roberto Aloisio in our joint work
on diffusion of cosmic rays, which is partly related to this work,
too.  Alexander Dolgov and Alexander Vilenkin are thanked for
valuable discussions. We are grateful to the anonymous referee for 
the valuable remark. We thank ILIAS-TARI for access to the LNGS
research infrastructure and for the financial support through EU
contract RII3-CT-2004-506222.


\begin{thebibliography}{}

\bibitem[Aloisio \& Berezinsky(2004)]{AloisioBere04} Aloisio, R., \& Berezinsky, V.
2005, \apj, 625, 249; 2004, astro-ph/0412578

\bibitem[Berezinsky \& Grigorieva(1988)]{Bere88}
Berezinsky, V.S., \& Grigorieva, S.I. 1988, \aap, 199, 1

\bibitem[Berezinsky et al.(1990a)]{book}
Berezinsky, V.S., Bulanov, S.V., Dogiel, V.A., Ginzburg, V.L., and
Ptuskin, V.S. 1990a, Astrophysics of Cosmic Rays, North-Holland

\bibitem[Berezinsky et al.(1990b)]{Bere90b}
Berezinsky, V.S., Dogiel, V.A., and Grigorieva, S.I. 1990b, \aap,
232, 582

\bibitem[Berezinsky et al.(2002)]{Bere02}
Berezinsky, V., Gazizov A.Z., and Grigorieva, S.I. 2002,
hep-ph/0204357; 2002, astro-ph/0210095

\bibitem[Kolb \& Turner(1990)]{KT} Kolb, E.W., \& Turner, M.S. 1990, 
The Early Universe, Westview Press

\bibitem[Lemoine(2005)]{Lemoine}
Lemoine, M. 2005, \prd, 71, 083007 (astro-ph/0411174)

\bibitem[Lifshitz \& Pitaevskii(2001)]{Lifshi} Lifshitz, E.M., \& 
Pitaevskii, L.P.
2001, Physical Kinetics (Vol.\ 10 of Theoretical Physics by
Landau, L.D., \& Lifshitz, E.M.) Fizmatlit

\bibitem[Peebles(1980)]{Peebles}
Peebles P.J.E. 1980, The Large-Scale Structure of the Universe,
Princeton Series in Physics

\bibitem[Syrovatskii(1959)]{Syrov}
Syrovatskii, S.I. 1959, Sov.\ Astron.\ J., 3, 22 [1959, Astron.\
Zh., 36, 17]

\bibitem[Weinberg(1972)]{Weinberg}
Weinberg, S. 1972, Gravitation and Cosmology, John Wiley and Sons

\end{thebibliography}
\end{document}